\documentclass[10pt,journal,compsoc]{IEEEtran}

\ifCLASSOPTIONcompsoc
  \usepackage[nocompress]{cite}
\else
  \usepackage{cite}
\fi

\usepackage{float}
\usepackage{cite}
\usepackage{amsmath,amssymb,amsfonts}
\usepackage{algorithm}
\usepackage{algorithmic}
\usepackage{graphicx}
\usepackage{textcomp}
\usepackage{xcolor}
\usepackage{multirow}
\usepackage{siunitx}
\usepackage{booktabs}
\usepackage{subfigure}
\usepackage{hyperref}
\usepackage{ragged2e}
\usepackage{threeparttable}
\usepackage{tikz}
\usepackage{enumitem}
\usepackage{xspace}
\usepackage{caption}
\usepackage{wrapfig}
\usepackage{graphicx}
\usepackage{enumitem}
\usepackage{amsmath}
\usepackage{makecell}
\usepackage{threeparttable}
\usepackage{algorithm}
\usepackage{algorithmic}
\usepackage{multirow}
\usepackage{diagbox}
\usepackage{makecell}
\usepackage{amsmath}
\usepackage{amssymb}
\usepackage{booktabs}

\def\BibTeX{{\rm B\kern-.05em{\sc i\kern-.025em b}\kern-.08em
    T\kern-.1667em\lower.7ex\hbox{E}\kern-.125emX}}

\begin{document}

\title{DDFAD: Dataset Distillation Framework for \\ Audio Data}

\author{Wenbo~Jiang,~\IEEEmembership{Member,~IEEE,}
        Rui~Zhang,~\IEEEmembership{Student Member,~IEEE,}
        Hongwei~Li (Corresponding author),~\IEEEmembership{Fellow,~IEEE,}
        Xiaoyuan~Liu,~\IEEEmembership{Student Member,~IEEE,}
        Haomiao~Yang,~\IEEEmembership{Senior Member,~IEEE,}
        and~Shui~Yu,~\IEEEmembership{Fellow,~IEEE}
\IEEEcompsocitemizethanks{

\IEEEcompsocthanksitem W. Jiang, R. Zhang, H. Li, X. Liu and H. Yang are with the School of Computer Science and Engineering, University of Electronic Science and Technology of China, China (e-mail: wenbo\_jiang@uestc.edu.cn, 202321081415@std.uestc.edu.cn, hongweili@uestc.edu.cn, xiaoyuan.l@foxmail.com, haomyang@uestc.edu.cn).
\IEEEcompsocthanksitem  S. Yu is a Professor of the School of Computer Science in the Faculty of Engineering and Information Technology at University of Technology Sydney, Sydney, Australia (e-mail: Shui.Yu@uts.edu.au).
}}



\IEEEtitleabstractindextext{%
\justify
\begin{abstract}
Deep neural networks (DNNs) have achieved significant success in numerous applications. The remarkable performance of DNNs is largely attributed to the availability of massive, high-quality training datasets. However, processing such massive training data requires huge computational and storage resources. Dataset distillation is a promising solution to this problem, offering the capability to compress a large dataset into a smaller distilled dataset. The model trained on the distilled dataset can achieve comparable performance to the model trained on the whole dataset.

While dataset distillation has been demonstrated in image data, none have explored dataset distillation for audio data. In this work, for the first time, we propose a Dataset Distillation Framework for Audio Data (DDFAD). Specifically, we first propose the Fused Differential MFCC (FD-MFCC) as extracted features for audio data. After that, the FD-MFCC is distilled through the matching training trajectory distillation method. Finally, we propose an audio signal reconstruction algorithm based on the Griffin-Lim Algorithm to reconstruct the audio signal from the distilled FD-MFCC. Extensive experiments demonstrate the effectiveness of DDFAD on various audio datasets. In addition, we show that DDFAD has promising application prospects in many applications, such as continual learning and neural architecture search.
\end{abstract}

}

\maketitle

\section{Introduction}
\begin{figure}[ht]
\centering
\includegraphics[width=0.48\textwidth]{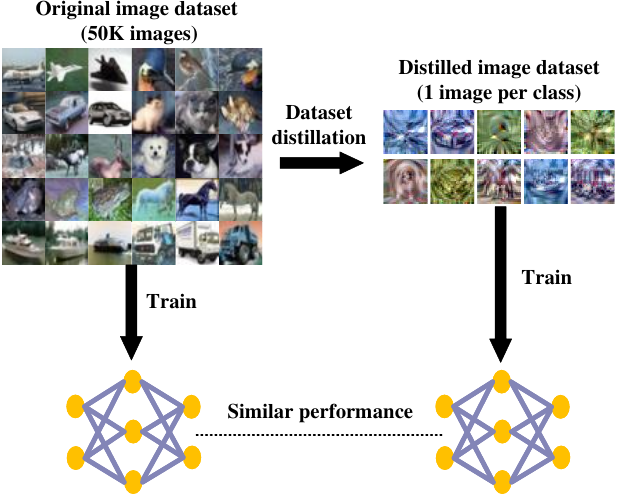}
\caption{Dataset distillation for image data.}
\label{Fig:Dataset distillation for image data}
\end{figure}

\IEEEPARstart{D}{NNs} have achieved remarkable performance in a wide range of applications. The superior performance of DNNs often requires large-scale training datasets. For example, the commonly used image classification dataset ImageNet \cite{ILSVRC15} has about 1.2 million training samples and 100,000 testing samples,  taking up about 148G of storage space; the commonly used object detection dataset COCO \cite{lin2014microsoft} has 118,287 training samples and 40,670 testing samples, taking up about 44G of storage space. However, the management of such massive data entails significant challenges, including the collection, storage, transmission, and pre-processing. Moreover, training models on such massive data comes with huge computational overhead. Such storage and computing requirements are huge and impractical for personal users. To address this problem, an emerging technology known as dataset distillation has garnered considerable research attention in recent years. As illustrated in Figure \ref{Fig:Dataset distillation for image data}, dataset distillation extracts the knowledge from a large-scale dataset and generates a smaller synthetic distilled dataset. The models trained on the distilled dataset can achieve performance comparable to those trained on the original dataset. This technique can significantly decrease the computational resources of training a DNN.  


\begin{figure}[ht]
\centering
\includegraphics[width=0.48\textwidth]{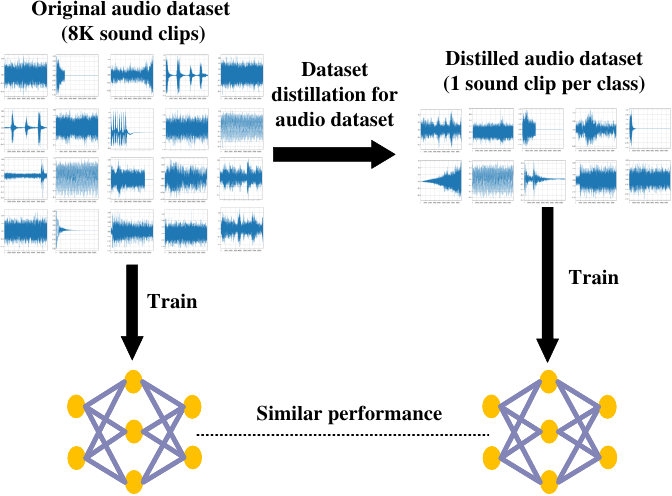}
\caption{Dataset distillation for audio data.}
\label{Fig:Dataset distillation for audio data}
\end{figure}

Existing works of dataset distillation are all focused on distilling image data, none of them has investigated dataset distillation for audio data. In fact, similar to the image dataset, the audio dataset also suffers from excessive volume, which needs huge storage and computational resources. For instance, the AudioSet dataset \cite{gemmeke2017audio} contains about 2 million audio clips from YouTube videos; the UrbanSound-8K dataset \cite{salamon2014dataset} includes 8,732 audio clips of 10 categories of urban sounds, taking up about 5.6G of storage space; There is also an urgent need for dataset distillation schemes for audio data. In this work, for the first time, we propose the Dataset Distillation Framework for Audio Data (DDFAD) for audio classification tasks. As illustrated in Figure \ref{Fig:Dataset distillation for audio data}, similar with dataset distillation for image data, DDFAD can compress a large number of audio clips into a smaller number of audio clips. Training with the distilled audio dataset can also achieve comparable performance to those trained on the original audio data.


However, it is non-trivial to distill audio data. While traditional feature extraction methods, such as linear predictive cepstral coefficient (LPCC) \cite{chowdhury2019fusing} and Mel frequency cepstral coefficient (MFCC) \cite{toffa2020environmental}, can extract audio data as feature spectrograms for DNN training, they prove insufficient in providing discriminative features in the case of the small-scale distilled dataset. To address this limitation, we propose the Fused Differential MFCC (FD-MFCC). It fuses the features of MFCC, the first-order difference of MFCC and the second-order difference of MFCC to make the extracted features more informative. After feature extraction, the dataset of FD-MFCC is distilled through the matching training trajectory (MTT) distillation method \cite{cazenavette2022dataset}\footnote{It is worth noting that other distillation methods are also applicable to DDFAD. In our experiments, we also consider other state-of-the-art distillation methods for evaluations.}. Finally, we propose an audio signal reconstruction algorithm based on the Griffin-Lim Algorithm (GLA) \cite{10354459} that rebuilds the distilled audio signal from the distilled FD-MFCC. 


In practice, DDFAD offers a method to compress large-scale audio dataset and is conductive to training audio data classification models and other downstream tasks such as continual learning and neural architecture search. The contributions of this work are summarized as follows:
\begin{itemize}
    \item We present DDFAD, the first dataset distillation framework for audio data. Specifically, we propose FD-MFCC to make the extracted features of the audio data more informative. After that, the dataset of FD-MFCC is distilled through the MTT distillation method. Finally, we propose an audio signal reconstruction algorithm based on GLA to reconstruct the audio signal from the distilled FD-MFCC.
    \item We conduct extensive evaluations on various audio datasets and DNN architectures to illustrate the effectiveness and cross-architecture generalization of DDFAD. 
    \item We carry out experiments to show that DDFAD can greatly improve various downstream applications, such as continual learning and neural architecture search.
\end{itemize}

The remainder of this paper is organized as follows: related work is presented in Section \ref{sec:Related Work}. Section \ref{sec:Methodology} provides the details of our attack methodologies. Experimental evaluations are shown in Section \ref{sec:Evaluation}. Potential applications are discussed in Section \ref{sec:Potential Applications}. Finally, Section \ref{sec:Conclusions} concludes the paper.

\section{Related Work}
\label{sec:Related Work}

\begin{figure*}
\centering
\includegraphics[width=0.97\textwidth]{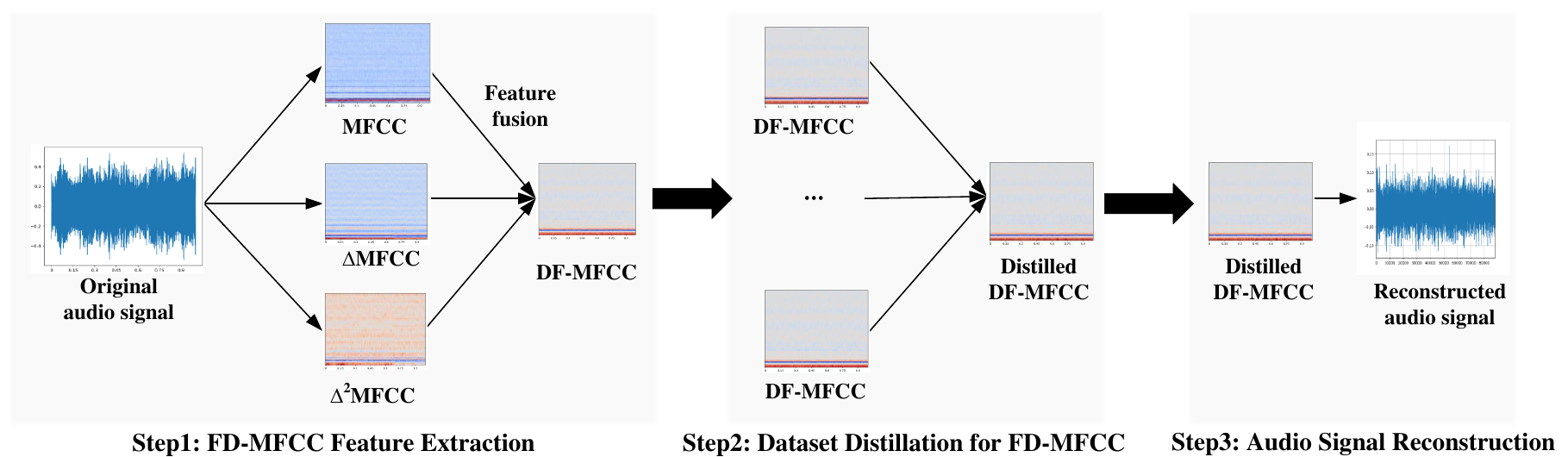}
\caption{The workflow of the proposed dataset distillation framework for audio data.}
\label{Fig:The workflow of DDFAD}
\end{figure*}

\subsection{Audio Data Classification}
\label{sec:Audio Data Classification}

Audio data classification is an important topic with many potential applications, such as speech classification in industrial automation, environmental sound classification in weather prediction, etc. Early works on audio classification used Support Vector Machines (SVM) \cite{lin2005audio}, K-Nearest Neighbors (KNN) \cite{lu2002content} and Hidden Markov Models (HMMs) \cite{pikrakis2006classification} for classification. More recently, audio classification based on DNN is gradually becoming mainstream and achieving leading performance \cite{li2022explainable,hasan2022genetic,ghosh2022automated,tripathi2023divide}. Thus, we mainly focus on DNN-based audio classification methods in this work.

In addition to choosing different machine learning models for classification, the extraction of audio features is also critical for audio data classification. Concretely, audio features can broadly be categorized into time-domain, frequency-domain, and cepstral-domain features. 

\textbf{Time-domain feature} characterizes audio signals in relation to time. It directly uses the one-dimensional (1D) audio signals as the input of the classification model, which is computation-efficient. Notable time-domain features include Short-Time Energy (STE) \cite{li2023precise}, Zero Crossing Rate (ZCR) \cite{bachu2010voiced}, Short-Time Autocorrelation Function (STAF) \cite{tang2023comparative}, etc.

\textbf{Frequency-domain feature} refers to the characteristics of audio signals varying in frequency. It no longer uses the original 1D signals, but 2D signals (i.e., spectrograms) as the model input. Therefore, frequency-domain features contain more information compared to time-domain features. Common frequency domain features include Fourier Transform (FT) \cite{shi2020novel}, Discrete Cosine Transform (DCT) \cite{geng2020end}, etc.

\textbf{Cepstral-domain feature} is mostly obtained by inverting some frequency-domain signals and their variants. Currently, the cepstral-domain feature is the most commonly used feature type and achieves leading performance in DNN-based audio classification. Representative cepstral-domain features include linear predictive cepstral coefficient (LPCC) \cite{chowdhury2019fusing}, Mel frequency cepstral coefficient (MFCC) \cite{toffa2020environmental}, etc.


\subsection{Dataset Distillation}
\label{sec:Dataset Distillation}
The concept of dataset distillation is originated from knowledge distillation \cite{hinton2015distilling}. Knowledge distillation is designed to transfer the knowledge of a large-scale model to a more lightweight model, whereas dataset distillation is a technique to distill the knowledge of the large-scale dataset a small-scale distilled dataset. Their underlying technologies are different.


In recent years, various dataset distillation algorithms have been proposed and they can be mainly categorized into three approaches: performance matching, parameter matching, and distribution matching.

\textbf{Performance matching.} This approach focuses on optimizing a distilled dataset to ensure that neural networks trained on it exhibit minimal loss on the original dataset, thereby achieving comparable performance between models trained on distilled and original datasets. Performance matching was first proposed by Wang et al. \cite{wang2018dataset}, who formulated dataset distillation as a bi-level optimization problem. However, the inner loops in their approach require extensive backpropagation gradient computation, which is highly inefficient. Subsequent works \cite{zhou2022dataset,nguyen2020dataset,nguyen2021dataset} proposed to replace the neural network in the inner loop with a kernel model, bypassing the backpropagation gradient computation process.

\textbf{Parameter matching.} This approach focuses on optimizing the consistency of trained model parameters between the distilled dataset and the original dataset. It was initially proposed by \textit{Zhao et al.} \cite{zhao2020dataset}, who formulated the objective as a minimization problem between two sets of gradients of the network parameters. Following \cite{zhao2020dataset}, numerous algorithms \cite{lee2022dataset,jiang2023delving,kim2022dataset,zhang2023accelerating,cazenavette2022dataset} have been proposed to improve parameter matching. For example, \textit{Cazenavette et al.} \cite{cazenavette2022dataset} proposed a multi-step parameter matching approach known as matching training trajectory (MTT); \textit{Zhao et al.} \cite{zhang2023accelerating} employed model augmentation techniques, such as utilizing early-stage models and parameter perturbation, to accelerate the training speed of dataset distillation.

\textbf{Distribution matching.} Instead of matching training effects or model parameters, distribution matching focuses on obtaining a distilled dataset whose distribution closely approximates that of the original dataset. For instance, \textit{Zhao et al.} \cite{zhao2023dataset} utilized the metric of Maximum Mean Discrepancy (MMD) metric to optimize the distance between the distribution of the distilled dataset and the original dataset; \textit{Wang et al.} \cite{wang2022cafe} proposed CAFE, which ensures that statistics of features for the distilled and original samples extracted by each network layer except the final one are consistent.


\subsection{Coreset Selection}
\label{sec:Coreset Selection}
Coreset selection is another strategy to select a representative subset of the whole dataset through heuristic selection criteria. It is commonly used in active learning to identify training samples. For instance, random selection \cite{rebuffi2017icarl} picks samples arbitrarily; Herding selection \cite{castro2018end} chooses samples closest to each class center; In K-Center selection \cite{chembu2023scalable}, multiple center points are chosen for each class in order to minimize the maximum distance between data points and their nearest center point. However, these coreset selection methods may not always yield optimal results for downstream audio data classification task. Furthermore, identifying an informative coreset can be challenging, particularly when dataset information is not concentrated in a few samples. In contrast, the dataset distillation methods can achieves better result for downstream audio data classification task through generating synthetic distilled data.

\section{Methodology}
\label{sec:Methodology}

\subsection{Overview}
In this section, we present the details of DDFAD. It mainly contains three phases: feature extraction for audio data, dataset distillation for FD-MFCC and audio signal reconstruction. The workflow of DDFAD is illustrated in Figure \ref{Fig:The workflow of DDFAD}. Below we describe the details of each phase.


\subsection{FD-MFCC Feature Extraction}
\label{sec:FD-MFCC Feature Extraction}
Currently, MFCC is the most commonly used feature type for audio data and achieves leading performance in DNN-based audio classification. Although MFCC feature performs well on the entire source dataset, in the case of dataset distillation, MFCC struggles to extract sufficient discriminative features to maintain the accuracy of the model trained on the distilled dataset. To overcome this limitation, we propose the Fused Differential MFCC (FD-MFCC), which fuses the features of MFCC, the first-order difference of MFCC and the second-order difference of MFCC. Since different difference orders of MFCC can represent features from different aspects, FD-MFCC can make full use of the complementarity between them and enhance the feature representation capability.



\begin{figure}
\centering
\includegraphics[width=0.48\textwidth]{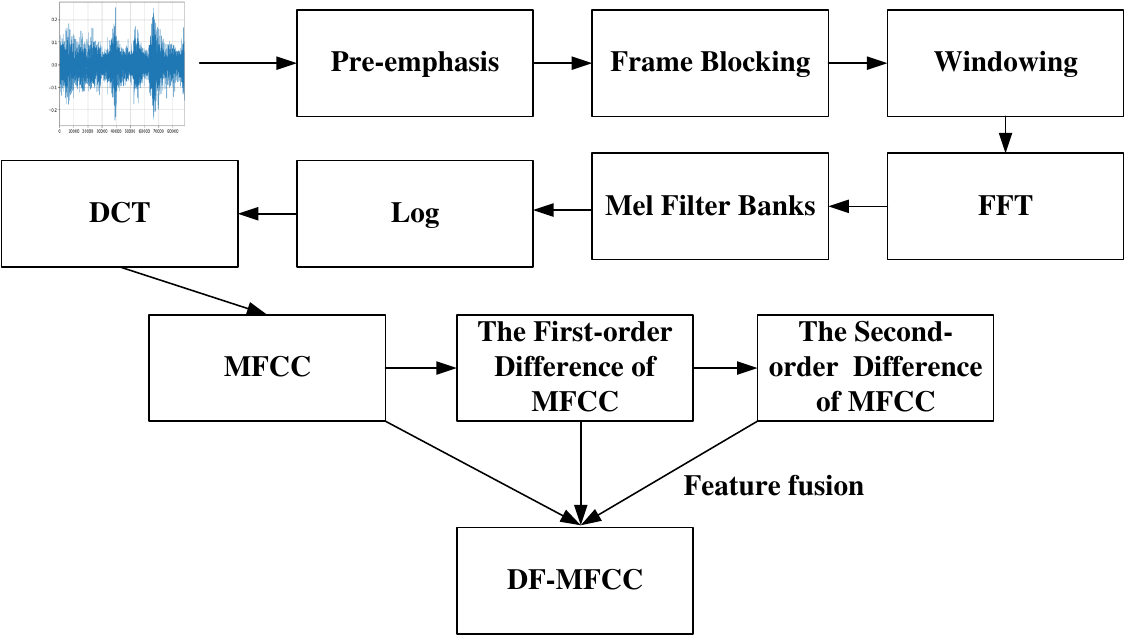}
\caption{The feature extraction process of FD-MFCC.}
\label{Fig:The process of feature extraction for audio data}
\end{figure}

As illustrated in Figure \ref{Fig:The process of feature extraction for audio data}, the feature extraction process of FD-MFCC consists of 9 steps:
\begin{enumerate}[leftmargin=*,itemsep=0pt,topsep=0pt]
    \item \textbf{Pre-emphasis.} This step compensates for the loss of the high-frequency part of the audio signal, which is beneficial for feature extraction.
    \item \textbf{Frame Blocking.} This step splits the audio data into small segments with a frame length of 20ms in order to ensure the smoothness of the audio signal.
    \item \textbf{Windowing.} This step enhances the strength of the middle part of the signal in each frame and weakens the discontinuities at the endpoints.
    \item \textbf{FFT.} FFT can transform time-domain signals into frequency-domain signals, which is conducive to obtaining more information in audio data.
    \item \textbf{Mel Filter Banks.} Since human ears are not sensitive to the low-frequency part of the audio signal, the Mel filter bank is designed to enhance the signal in the middle of the triangle wave region and weaken the signal on both sides, which is beneficial for feature extraction.
    \item \textbf{Log.} Since the perception of human ears grows in a logarithmic way, it also needs to take the logarithm of the obtained features to simulate the perception of human ears.
    \item \textbf{DCT.} The function of DCT is to remove the correlation between the signals of different orders and map the signals back into a lower dimensional space. The MFCC feature can be obtained by performing DCT on the Mel spectrum feature.
    \item \textbf{The first-order and second-order difference of MFCC.} After obtaining MFCC, we further calculate the first-order and second-order difference of MFCC to extract more features:
\begin{equation}
\label{eq:the first-order difference of MFCC}
\Delta MFCC(t)=\frac{MFCC(t+1)-MFCC(t)}{2},
\end{equation}
\begin{equation}
\label{eq:the second-order difference of MFCC}
\Delta^2 MFCC(t)=\frac{\Delta MFCC(t+1)-\Delta MFCC(t)}{2},
\end{equation}
        where MFCC(t) denotes the value of MFCC at time t; $\Delta$MFCC and  $\Delta^2$MFCC denote the first-order and second-order difference of MFCC, respectively.
    \item \textbf{Feature Fusion.} The MFCC, $\Delta$MFCC and $\Delta^2$MFCC are spliced and fused together to get FD-MFCC.
\end{enumerate}


In our experiments, we conduct extensive ablation studies to show the superiority of our proposed FD-MFCC compared with traditional MFCC (refer to Section \ref{sec:Ablation Study of FD-MFCC} for more details).

\subsection{Dataset Distillation for FD-MFCC}
\label{sec:Dataset Distillation for MFCC}

Given the source training dataset $\mathcal{S}$ with $|\mathcal{S}|$ samples, the objective of dataset distillation is to extract the knowledge of $\mathcal{S}$ into a small distilled dataset $ \mathcal{D}$ with $|\mathcal{D}|$ samples ($|\mathcal{S}|\gg|\mathcal{D}|$), and the model trained on $\mathcal{D}$ can achieve comparable performance to the model trained on $\mathcal{S}$. 

In this work, we adopt the state-of-the-art (SOTA) matching training trajectory (MTT) distillation method \cite{cazenavette2022dataset} in DDFAD to distill FD-MFCC. Algorithm \ref{alg:The Process of Dataset Distillation} illustrates the detailed process of dataset distillation for FD-MFCC. Specifically, MTT first trains models on $\mathcal{S}$ and collects the trajectories of the model (referred to as the teacher model) in the buffer. Subsequently, ingredients in the buffer are randomly chosen to initialize the student model (the model trained on $\mathcal{D}$). After collecting the trajectories of the student model, the distilled dataset is updated by matching the two training trajectories. The objective loss of MTT is defined as:
\begin{equation}
\label{eq:loss of trajectory matching}
\mathcal{L}_o=\frac{\left\|\boldsymbol{\theta}_{\mathcal{D}}^{(t+N)}-\boldsymbol{\theta}_{\mathcal{S}}^{(t+M)}\right\|_2^2}{\left\|\boldsymbol{\theta}_{\mathcal{S}}^{(t)}-\boldsymbol{\theta}_{\mathcal{S}}^{(t+M)}\right\|_2^2},
\end{equation}
where $\boldsymbol{\theta}_{\mathcal{S}}^{(t)}$ represents the parameter of the teacher model at training epoch $t$, which is stored in the buffer; $\boldsymbol{\theta}_{\mathcal{D}}^{(t+N)}$ represents the parameter of the student model trained on $\mathcal{D}$ for $N$ epochs with the initialization of $\boldsymbol{\theta}_{\mathcal{S}}^{(t)}$.


\begin{algorithm}
\caption{Dataset Distillation for FD-MFCC}
\label{alg:The Process of Dataset Distillation}
\begin{algorithmic}[1]
\REQUIRE $\{\boldsymbol{\theta}_{\mathcal{S}}^{(i)}\}_{i=1}^T$: trajectories of the teacher model; $M$: number of iterations for dataset distillation; $N$: number of updates between starting and target expert parameters; $T'$: the maximum start epoch. 
\ENSURE the distilled dataset $\mathcal{D}$ and the learning rate $\alpha$.
\STATE randomly initialize the distilled dataset $\mathcal{D}$ and the trainable learning rate $\alpha$
\FOR{$m = 0 \rightarrow M$}
\STATE Choose random start epoch $t<T'$
\STATE Initialize student network with teacher trajectories: $\boldsymbol{\theta}_{\mathcal{D}}^{(t)}=\boldsymbol{\theta}_{\mathcal{S}}^{(t)}$
\FOR {$n = 0 \rightarrow N$}
\STATE Sample a batch of distilled data: $b_{t+n} \in \mathcal{D}$
\STATE Train the student model with gradient descent method: $\boldsymbol{\theta}_{\mathcal{D}}^{(t+n+1)}=\boldsymbol{\theta}_{\mathcal{D}}^{(t+n)}-\alpha\nabla \mathcal{L}(b_{t+n};\boldsymbol{\theta}_{\mathcal{D}}^{(t+n)})$
\ENDFOR
\STATE Compute the objective $\mathcal{L}_o$ according to Eq.(\ref{eq:loss of trajectory matching})
\STATE Update $\mathcal{D}$ and $\alpha$ w.r.t. $\mathcal{L}_o$
\ENDFOR
\RETURN  $\mathcal{D}$ and $\alpha$
\end{algorithmic}
\end{algorithm}

\subsection{Audio Signal Reconstruction from the Distilled FD-MFCC}
\label{sec:Restore Distilled Features Data to Audio Data}
After distilling the dataset of FD-MFCC, the final phase is to reconstruct the audio signal from the distilled FD-MFCC. We propose an audio signal reconstruction algorithm based on GLA to reconstruct the audio signal from the distilled FD-MFCC.

As presented in Algorithm \ref{alg:The Process of Audio Signal Reconstruction from the Distilled FD-MFCC}, we begin by applying the inverse DCT (IDCT) to the distilled FD-MFCC to obtain the dB-scaled spectrogram. Subsequently, we employ the $\operatorname{dB-to-power}$ function\footnote{Available in librosa package: https://github.com/librosa/librosa. \label{librosa}} to map the dB-scaled spectrogram to the mel power spectrogram. After that, we use the $\operatorname{mel-to-stft}$ function\textsuperscript{\ref {librosa}} to approximate Short-Time Fourier Transform (STFT) magnitude from a Mel power spectrogram. Finally, we employ GLA \cite{10354459} to reconstruct the audio signal from the STFT magnitude. 


Concretely, the objective of GLA is to reconstruct a spectrogram that is consistent with the given amplitude $\mathbf{A}$. This is achieved through the following alternative projection procedure:
\begin{equation}
\label{eq:GLA}
\mathbf{X}^{[i]}=P_{\mathcal{C}}\left(P_{\mathcal{A}}\left(\mathbf{X}^{[i-1]}\right)\right),
\end{equation}
where $\mathbf{X}^{[i]}$ denotes the reconstructed audio signal at the $i$th iteration, $\mathcal{C}$ is defined as the set of all spectrograms
that corresponds to a time-domain signal, $\mathcal{A}$ is defined as the set of all spectrograms that have the given magnitude spectrogram $\mathbf{A}$. $P_{\mathcal{A,C}}$ represents the projection onto set $\mathcal{A,C}$ and they are defined as follows:
\begin{equation}
\label{eq:projections c}
P_{\mathcal{A}}(\mathbf{X})=\mathbf{A} \frac{\mathbf{X}}{|\mathbf{X}|},
\end{equation}
\begin{equation}
\label{eq:projections a}
P_{\mathcal{C}}(\mathbf{X})=\operatorname{STFT}(\operatorname{iSTFT}(\mathbf{X})),
\end{equation}
where $\operatorname{STFT}$ represents the short-time Fourier transform and $\operatorname{iSTFT}$ represents the inverse $\operatorname{STFT}$. The reconstruction process is executed iteratively for $I$ rounds to obtain the final reconstructed audio signal. The reconstruct distilled audio data can be seen in Figure \ref{Fig:Visualizations of the Distilled Audio Data}.

\begin{algorithm}
\caption{Audio Signal Reconstruction from the Distilled FD-MFCC}
\label{alg:The Process of Audio Signal Reconstruction from the Distilled FD-MFCC}
\begin{algorithmic}[1]
\REQUIRE $\mathcal{D}$: the distilled dataset of FD-MFCC; $I$: the maximum number of iteration.
\ENSURE the reconstructed audio signal.
\STATE Compute dB-scaled spectrograms: \newline $Spectrogram_{dB} = \operatorname{IDCT}(\mathcal{D})$
\STATE Compute mel power spectrograms: \newline $Spectrogram_{mel} = \operatorname{dB-to-power}(Spectrogram_{dB})$
\STATE Compute STFT magnitude: \newline $\mathbf{A} = \operatorname{mel-to-stft}(Spectrogram_{mel})$
\STATE Randomly initialize $\mathbf{X}^{[0]}$
\FOR{$i = 1 \rightarrow I$}
\STATE $\mathbf{X}^{[i]}=P_{\mathcal{C}}\left(P_{\mathcal{A}}\left(\mathbf{X}^{[i-1]}\right)\right)$
\ENDFOR
\RETURN $\mathbf{X}^{[i]}$
\end{algorithmic}
\end{algorithm}



\section{Evaluation}
\label{sec:Evaluation}

\subsection{Experimental Setup}
\label{sec:Experimental Setup}
\subsubsection{Model Architecture}
We consider three model architectures, including ResNet18 \cite{he2016deep}, ConvNet (which mainly contains multiple Conv-ReLU-AvgPooling blocks) and VGG11 \cite{simonyan2015very} for audio data classification. 

\subsubsection{Datasets}
\begin{itemize}
\item \textbf{Free Spoken Digit Dataset (FSDD) \cite{FSDD}.} FSDD comprises recordings of spoken digits at a sampling rate of 8kHz. It contains 3,000 audio data clips of English pronunciation of numbers (0-9) recorded by six speakers.
\item \textbf{UrbanSound-8k Dataset \cite{salamon2014dataset}.} The UrbanSound dataset includes 8,732 labeled audio clips, each lasting up to 4 seconds. It covers 10 categories of urban sounds, such as air conditioner, car horn, drilling, etc.
\item \textbf{Ryerson Audio-Visual Database of Emotional Speech and Song (RAVDESS) \cite{livingstone2018ryerson}.}  The RAVDESS dataset contains 7,356 audio clips depicting seven different speech emotions, e.g., calm, happy, sad, etc.
\end{itemize}

\subsubsection{Distillation Methods}
In our experiments, we consider three SOTA distillation methods for experimental evaluations. 
\begin{itemize}
\item \textbf{Dataset Condensation with Gradient Matching (DCGM) \cite{zhao2020dataset}.} DCGM aims to minimize the gap between two sets of gradients of the network parameters, where the gradients are computed based on the training loss over both the original dataset and the distilled dataset.
\item \textbf{Dataset Condensation with Differentiable Siamese Augmentation (DCDSA) \cite{zhao2021dataset}.} DCDSA proposes a differentiable siamese augmentation method to synthesize distilled data to obtain better performance.
\item \textbf{Matching Training Trajectory (MTT) \cite{cazenavette2022dataset}.} MTT first trains models on the source training dataset and collects the trajectories of the model in the buffer. After that, it also collects the trajectories of the network trained on the distilled dataset. The distilled dataset is updated by matching the two training trajectories.
\end{itemize}

Besides, we also consider two coreset selection methods for comparison. 
\begin{itemize}
\item \textbf{Random selection \cite{rebuffi2017icarl}.} This is a simple approach that randomly selects samples as the coreset.
\item \textbf{Herding selection \cite{castro2018end}.} This is a distance-based algorithm that selects samples whose center is close to the center of each class. 
\end{itemize}

\subsection{Distillation Performance}

\begin{table*}\small
\centering
\caption{The test accuracy (\%) of DDFAD under different dataset distillation methods.}
\label{table:The test accuracy under DDFAD}
\resizebox{0.98\linewidth}{!}{
\begin{threeparttable}
\begin{tabular}{ccc|ccc|cc|c}
  \toprule
  \multirow{2}{*}{Dataset} & \multirow{2}{*}{Architecture}&\multirow{2}{*}{CPC} & \multicolumn{3}{c|}{Distillation methods}  & \multicolumn{2}{c|}{Coreset selection methods} & \multirow{2}{*}{Whole dataset}\\
  \cline{4-8}
  &  &  &DCGM \cite{zhao2020dataset}  & DCDSA \cite{zhao2021dataset} &   MTT \cite{cazenavette2022dataset} &Random \cite{rebuffi2017icarl} &Herding \cite{castro2018end}&  \\
  \toprule
  \multirow{9}{*}{FSDD}&  \multirow{3}{*}{ResNet18}
  & 1 & 17.75	&17.86	  & \textbf{28.42} &  	11.57  & 14.31 & \multirow{3}{*}{98.73}\\
  &&10 & 58.97	&	56.05  & \textbf{75.30} &  42.98	  & 45.69 &\\
 &&50  & 87.46	&	94.12  & \textbf{97.30} &  80.78	  & 85.13 &\\
  \cline{2-9}
  &\multirow{3}{*}{ConvNet}
  & 1 & 14.32	&	17.01  & \textbf{29.88} &  11.02	  &13.20  &\multirow{3}{*}{90.52}\\
  &&10 & 55.75	&	54.33  & \textbf{65.60} &  42.66	  & 47.50 &\\
 &&50  & 	77.81&	80.29  & \textbf{89.67} &  69.79	  & 73.08 &\\
  \cline{2-9}
  &\multirow{3}{*}{VGG11}
  & 1 & 21.18	&	14.97  &\textbf{21.96}  &  11.97	  & 14.21 & \multirow{3}{*}{98.44}\\
  &&10 & 55.44	&	52.50  & \textbf{58.98} &  44.08	  & 48.27 &\\
 &&50  & 87.07	&	89.26  & \textbf{94.38} &  81.13	  & 85.59 &\\
  \hline  
    \multirow{9}{*}{UrbanSound}&  \multirow{3}{*}{ResNet18}
  & 1 & 12.48	&	12.04  & \textbf{19.88} &  10.84	  & 11.99 & \multirow{3}{*}{93.89}\\
  &&10 & 38.30	&	34.57  &  \textbf{40.69}&  21.80	  & 23.26 &\\
 &&50  & 53.81	&	52.58  & \textbf{62.75} &  43.77	  & 45.84 &\\
  \cline{2-9}
  &\multirow{3}{*}{ConvNet}
  & 1 & 11.30	&	12.55  & \textbf{21.59} &  10.35	  &  11.51& \multirow{3}{*}{70.78}\\
  &&10 & 30.08	&22.92	  &  \textbf{42.97}&  21.29	  &  22.08&\\
 &&50  & 53.56	&	55.19  &\textbf{68.22} &  42.62	  &  43.64&\\
  \cline{2-9}
  &\multirow{3}{*}{VGG11}
  & 1 & 10.82	&	11.60  & \textbf{19.63} &  10.11	  &  11.37& \multirow{3}{*}{89.04}\\
  &&10 & 21.34	&17.55	  & \textbf{27.47} &  15.33	  &  16.88&\\
 &&50  & 42.65	&	44.32  & \textbf{59.10} &  33.29	  & 33.30 &\\
    \hline  
    \multirow{9}{*}{RAVDESS }&  \multirow{3}{*}{ResNet18}
  & 1 & 12.54	&14.77	  & \textbf{17.30} &  11.52	  &12.63  & \multirow{3}{*}{67.75}\\
  &&10 & 23.28	&	25.98  & \textbf{32.50} &  18.38	  & 20.71 &\\
 &&50  & 36.60	&39.78	  &\textbf{48.96}  &  33.04	  & 34.12 &\\
  \cline{2-9}
  &\multirow{3}{*}{ConvNet}
  & 1 & 16.30	&	14.64  & \textbf{22.17} &  10.89	  & 12.72 & \multirow{3}{*}{49.81}\\
  &&10 & 20.83	&	21.51  &\textbf{29.22}  &  11.97	  &  13.65&\\
 &&50  & 33.70	&	 32.89 & \textbf{46.91} &  24.25	  & 25.88 &\\
  \cline{2-9}
  &\multirow{3}{*}{VGG11}
  & 1 & 13.55	&12.79	  & \textbf{19.20} &  11.22	  & 11.89 & \multirow{3}{*}{62.32}\\
  &&10 & 	26.85&	23.69  & \textbf{27.33} &  18.83	  &  19.07&\\
 &&50  & 	39.02&	38.45  & \textbf{47.91} &  34.46	  &  35.44&\\
    \bottomrule
\end{tabular}
    \end{threeparttable}}
\end{table*}

We perform our DDFAD with five considered distillation methods (DCGM \cite{zhao2020dataset}, DCDSA \cite{zhao2021dataset}, MTT \cite{cazenavette2022dataset}, Random selection \cite{rebuffi2017icarl}, Herding selection \cite{castro2018end}), to synthesize 1, 10, and 50 clips per class (CPC) respectively. 

As presented in Table \ref{table:The test accuracy under DDFAD}, dataset distillation methods are much more effective than coreset selection methods with the same CPC. This is because the dataset information can not be concentrated in a few samples. The synthetic distilled data can better represent the information of the whole dataset. Furthermore, among these dataset distillation methods, the DDFAD incorporating with the MTT outperforms other dataset distillation methods. Thus, for subsequent experiments, we adopt the MTT distillation method for evaluations of DDFAD.

For instance, in the case of the FSDD dataset with CPC=50, where the distilled dataset accounts for only 1/6 of the whole dataset, the test accuracies of DDFAD with MTT (97.30\% for ResNet18, 89.67\% for ConvNet and 94.38\% for VGG11) are very close to the test accuracies of the model trained on the whole dataset (98.73\% for ResNet18, 90.52\% for ConvNet and 98.44\% for VGG11). The storage space and computational resources required by the whole dataset is six times more than the storage space and computational resources required by the distilled dataset.

\subsection{Cross-architecture Generalization}
\label{sec:Cross-architecture Generalization}
In dataset distillation, it is crucial for the distilled dataset constructed on one model to yield similar training effects on downstream models with arbitrary architectures. Thus, in this subsection, we evaluate the cross-architecture generalization performance of DDFAD. Specifically, we utilize the distilled datasets constructed on ResNet18, ConvNet and VGG11 to train models with different architectures, including ResNet18, ConvNet, and VGG11. The results in Table \ref{Table: Cross-architecture generalization} show that the performance of DDFAD remains consistent across different models used for distillation, which demonstrates the good cross-architecture generalization of DDFAD.

\begin{table}[ht]\small
\begin{center}
\caption{Cross-architecture generalization of DDFAD (CPC=50).}
\label{Table: Cross-architecture generalization}
\resizebox{0.98\linewidth}{!}{
\begin{threeparttable}
\begin{tabular}{c|c|ccc}
\toprule
  \multirow{2}{*}{Dataset} & Distillation & \multicolumn{3}{c}{Evaluation architecture}  \\
    \cline{3-5}
  &  architecture & ResNet18 & ConvNet & VGG11 \\
\hline
\multirow{3}{*}{FSDD}& ResNet18 & 97.30 & 78.33  & 80.15    \\
&ConvNet  & 96.07 & 89.65  &   92.33  \\
&VGG11  & 96.31 & 85.89  &    94.38 \\
  \cline{1-5}
\multirow{3}{*}{UrbanSound}& ResNet18 & 62.75 & 56.79   &  51.88    \\
&ConvNet  & 62.06 &68.22   &  39.81   \\
&VGG11  &  66.33& 58.87  &  59.10   \\
  \cline{1-5}
\multirow{3}{*}{RAVDESS}& ResNet18 &48.96  & 47.11  &  48.35   \\
&ConvNet  &  47.54& 46.91  &   49.63  \\
&VGG11  & 47.28 &  47.10 &   47.91  \\
\bottomrule
\end{tabular}
    \end{threeparttable}}
\end{center}
\end{table}

\subsection{Ablation Study of FD-MFCC}
\label{sec:Ablation Study of FD-MFCC}
\begin{figure*}[ht]
\centering
\subfigure[FSDD] {\includegraphics[width=2in]{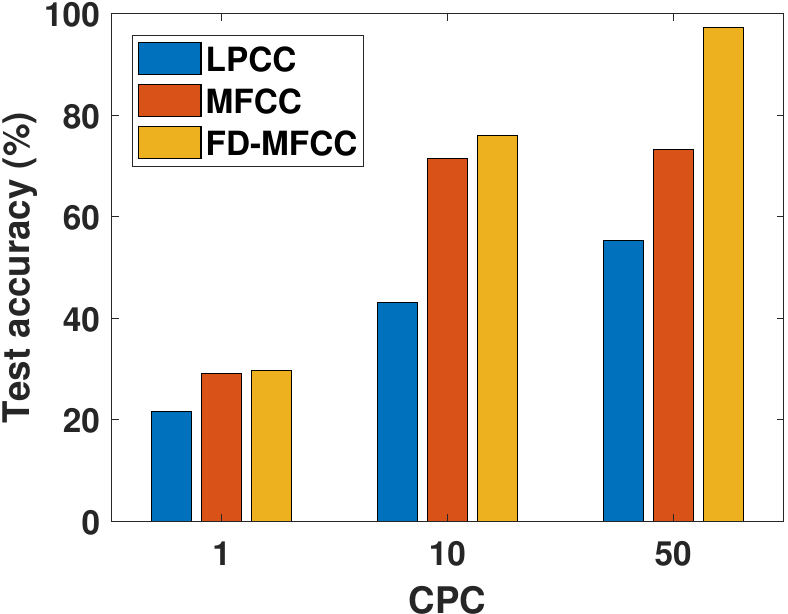}}
\hspace{5mm}
\subfigure[UrbanSound] {\includegraphics[width=2in]{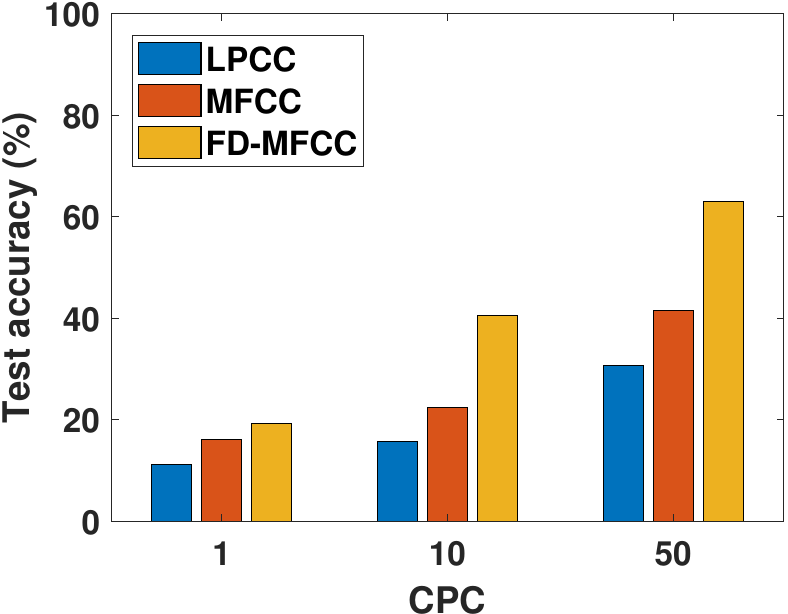}}
\hspace{5mm}
\subfigure[RAVDESS] {\includegraphics[width=2in]{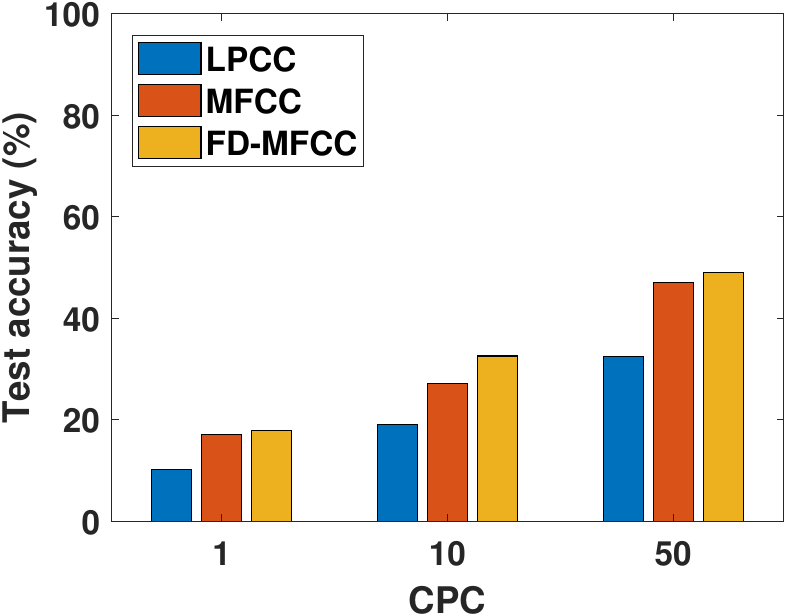}}
\caption{Ablation Study of FD-MFCC.}
\label{fig: Ablation Study of FD-MFCC}
\end{figure*}

In this subsection, we conduct ablation studies to compare the effectiveness of our proposed FD-MFCC with MFCC \cite{toffa2020environmental} and LPCC \cite{chowdhury2019fusing} in DDFAD. As depicted in Figure \ref{fig: Ablation Study of FD-MFCC}, for different datasets and CPC settings, FD-MFCC consistently outperforms traditional MFCC and LPCC. This superiority arises from the ability of FD-MFCC to leverage the complementary nature of different difference orders of MFCC, thereby enhancing feature representation capability. In scenarios where the number of training samples is limited, such as dataset distillation, FD-MFCC is more informative and therefore can achieve higher accuracy.

\subsection{Analysis of the Resources Requirement of DDFAD}
\label{sec:Computational Overhead}
\begin{table*}[ht]\small
\centering
\caption{The resources requirement of DDFAD.}
\label{table:The computational overhead of DDFAD}
\begin{threeparttable}
\begin{tabular}{cc|cccc|cccc}
  \toprule
  \multirow{2}{*}{Dataset} & \multirow{2}{*}{Architecture} & \multicolumn{4}{c|}{Distillation time (min)} & \multicolumn{4}{c}{GPU memory (GB)}\\
    \cline{3-10} 
  &  &CPC=1   & CPC=10 &   CPC=20 &CPC=50   &CPC=1   & CPC=10 &   CPC=20 &CPC=50 \\
  \toprule
  \multirow{3}{*}{FSDD}&  ResNet18 &  	80.67 &92.11 &100.59 &116.04 &31.01&	35.61&	38.45&	44.22 \\
  &  ConvNet &  12.31 &15.16 & 15.79 &	17.20& 17.79	&21.00	&23.29&	25.24 \\
 &  VGG11 & 50.21 & 52.84& 59.38& 70.81	&15.50&	20.59&	24.23&	29.81  \\
    \hline  
\multirow{3}{*}{UrbanSound}&  ResNet18 &  179.56	& 311.55 	 &316.80   & 337.91 &36.08	&46.71&	52.61	&60.07 \\
  &  ConvNet &  17.16	&  21.66	 &  23.82 &  29.57 &20.65	&25.90&	28.58&	34.36 \\
 &  VGG11 &  93.13	&  105.18	 & 108.20  & 11.89 & 18.31	&26.52&	36.92&	46.12\\
    \hline  
 \multirow{3}{*}{RAVDESS}&  ResNet18 &  105.11 &  	107.30 & 111.56  &  113.88 &37.94&	44.00	&57.18&	72.80 \\
  &  ConvNet &  18.33	&  21.19	 & 22.42  &  23.86	&19.49&	26.88	&30.19	&47.08\\
 &  VGG11 &  	83.51	& 85.35 	 &86.83   &  104.01	& 24.01&	43.09&	50.58&	56.91 \\
    \bottomrule
\end{tabular}
    \end{threeparttable}
\end{table*}

For dataset distillation, it is also important to consider the cost of resources of the algorithm. Hence, we report the computational overhead and the occupied GPU memory of DDFAD across different datasets, model architectures and CPC settings in Table \ref{table:The computational overhead of DDFAD}. All the experiments are run on NVIDIA RTX A6000 GPUs. 

The results indicate that more complex model architectures are often accompanied by larger computational overhead and more GPU memory. Overall, the time overhead and occupied GPU memory of DDFAD is also acceptable for data owners and can be further reduced on better devices.

\subsection{Visualizations of the Distilled Audio Data}
\label{sec:Visualizations of the Distilled Audio Data}
\begin{figure*}[ht]
\centering
\subfigure[Digital seven audio data from the FSDD dataset] {\includegraphics[width=0.47\textwidth]{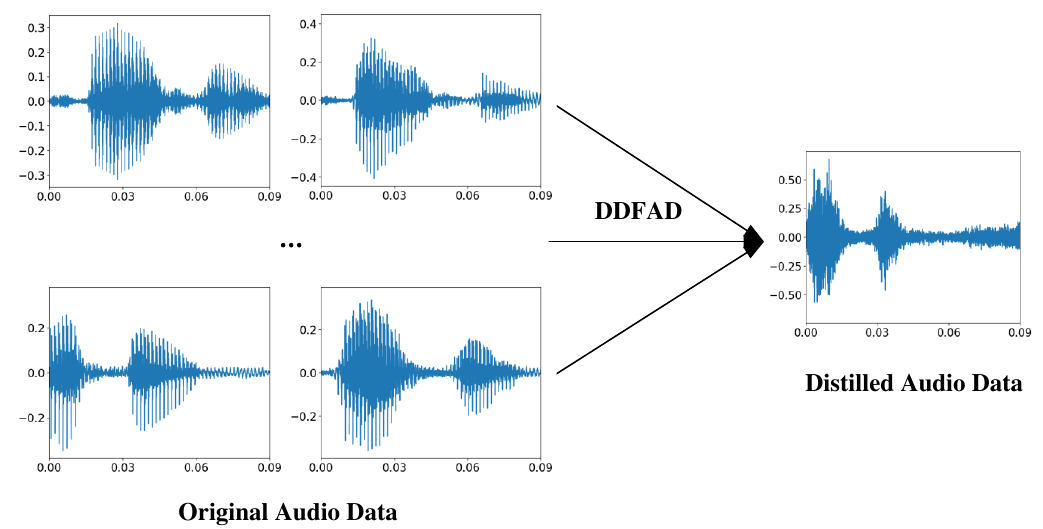}}
\hspace{5mm}
\subfigure[Dog bark audio data from the UrbanSound-8k dataset] {\includegraphics[width=0.47\textwidth]{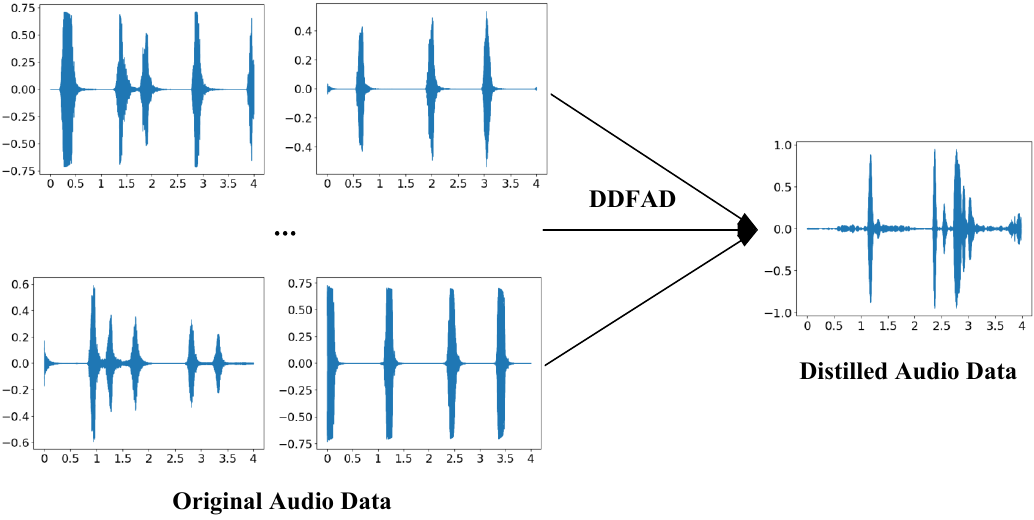}}
\subfigure[Gun shot audio data from the UrbanSound-8k dataset] {\includegraphics[width=0.47\textwidth]{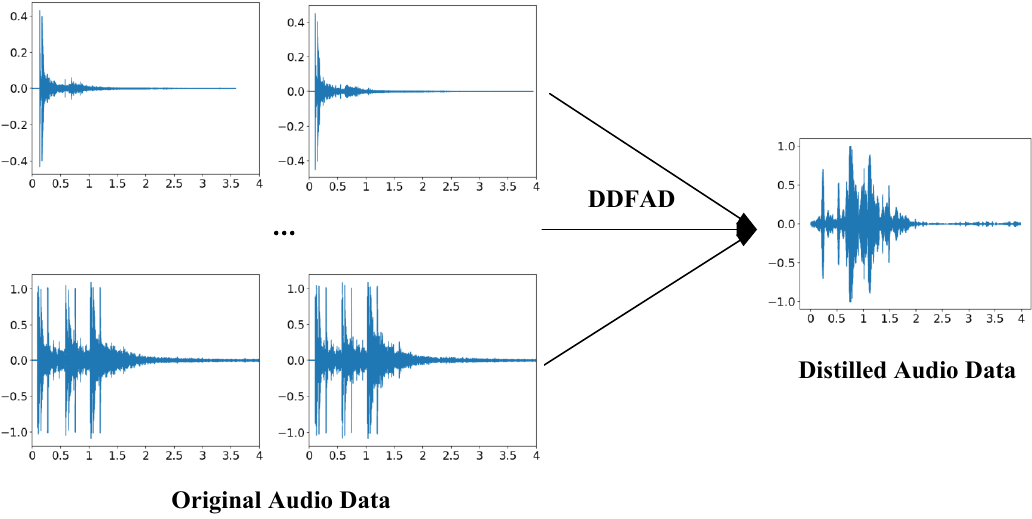}}
\hspace{5mm}
\subfigure[Calm speech emotions audio data from the RAVDESS dataset] {\includegraphics[width=0.47\textwidth]{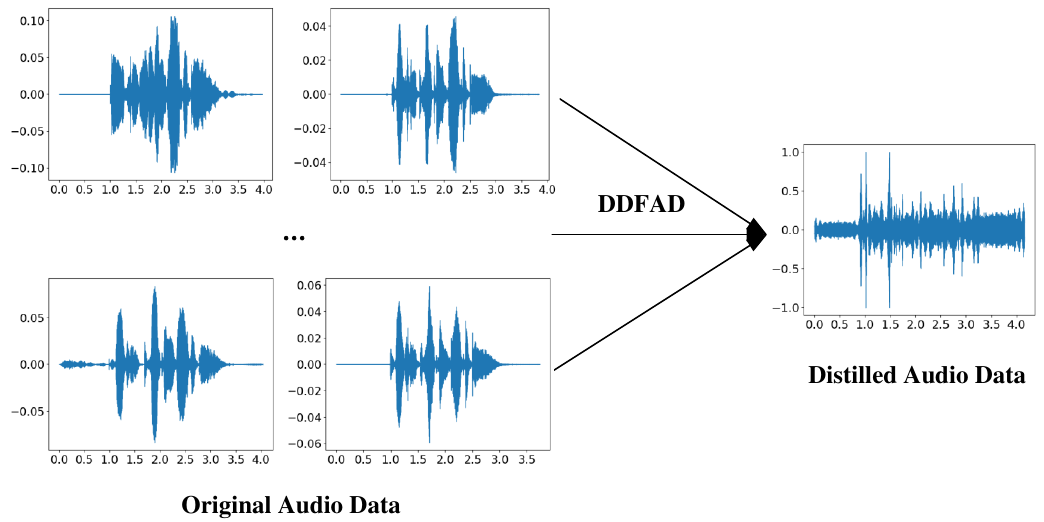}}
\caption{Visualizations of the recovered distilled audio data.}
\label{Fig:Visualizations of the Distilled Audio Data}
\end{figure*}

We select the digital seven audio data from the FSDD dataset; dog bark and gun shot audio data from the UrbanSound-8k dataset; and calm speech emotions audio data from the RAVDESS dataset as examples to show the waveform diagram of the original audio data and distilled audio data (recovered by Algorithm \ref{alg:The Process of Audio Signal Reconstruction from the Distilled FD-MFCC}). As presented in Figure \ref{Fig:Visualizations of the Distilled Audio Data}, similar with dataset distillation for image data (see Figure \ref{Fig:Dataset distillation for image data}), the distilled audio data may not closely resemble the original audio data. It may be synthetic audio data with no practical meaning, but is conductive to the subsequent training process of audio data classification tasks.

\subsection{Robustness against Noise}
\label{sec:Robustness against Noise}
\begin{figure*}[ht]
\centering
\subfigure[ResNet18] {\includegraphics[width=2in]{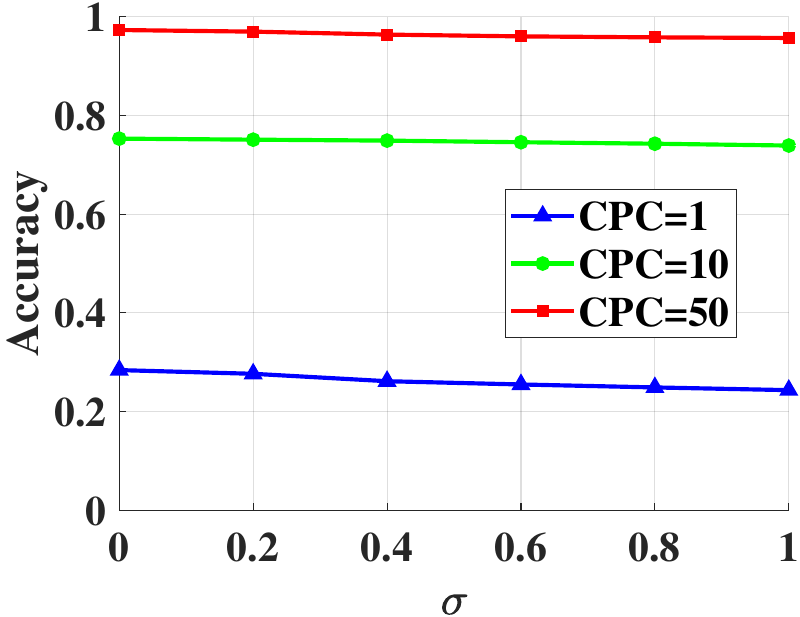}}
\hspace{5mm}
\subfigure[ConvNet] {\includegraphics[width=2in]{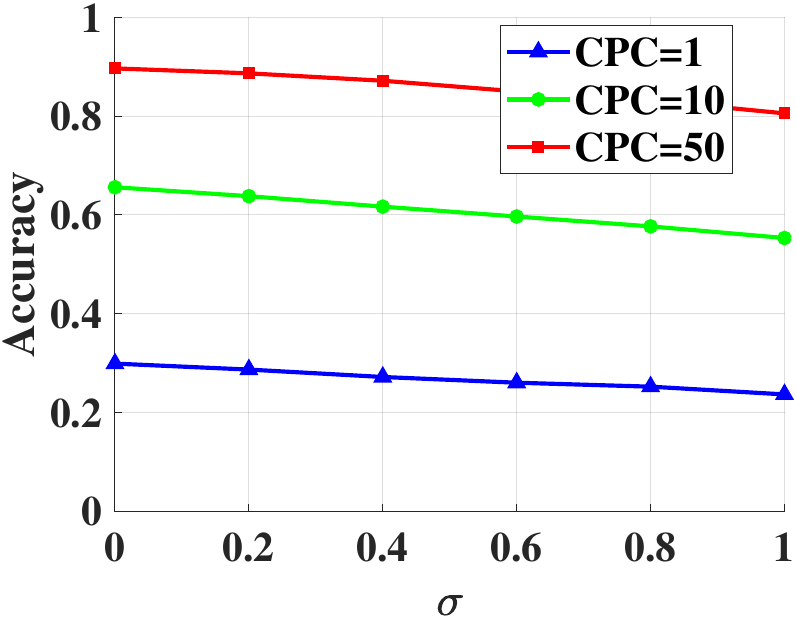}}
\hspace{5mm}
\subfigure[VGG11] {\includegraphics[width=2in]{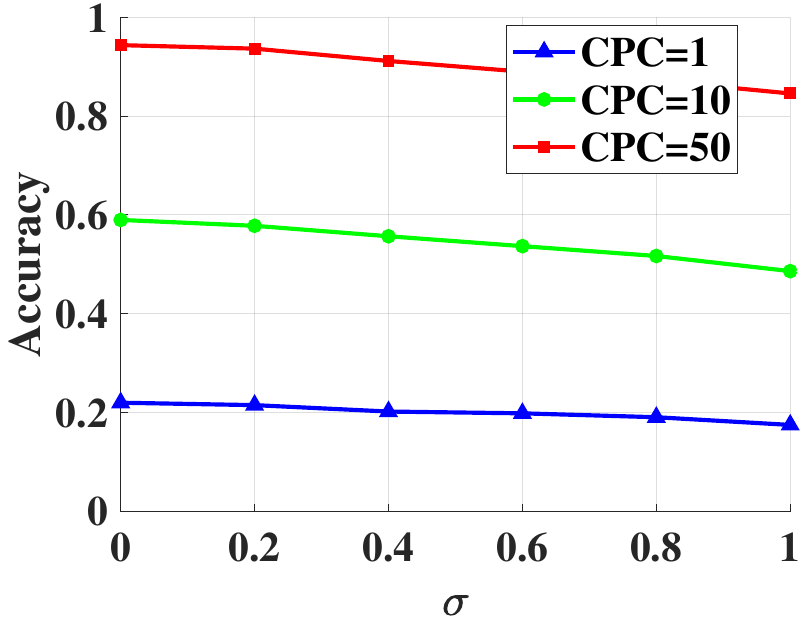}}
\caption{Robustness against Gaussian noise with different variance $\sigma$.}
\label{Fig:Robustness against Noise}
\end{figure*}

In this subsection, we evaluate the performance of DDFAD in the presence of additional noise. Specifically, we select the FSDD dataset as an example, and artificially add Gaussian noise with different variance $\sigma$ to the waveforms of the distilled data to train the classification model. As presented in Figure \ref{Fig:Robustness against Noise}, the accuracy remains stable with increasing Gaussian noise levels. It demonstrates that the distilled training dataset is robust to additional Gaussian noise. 

\section{Potential Applications}
\label{sec:Potential Applications}
In this section, we discuss some potential applications of DDFAD, such as continual learning and neural architecture search.

\begin{figure*}[ht]
\centering
\subfigure[DDFAD] {\includegraphics[width=1.75in]{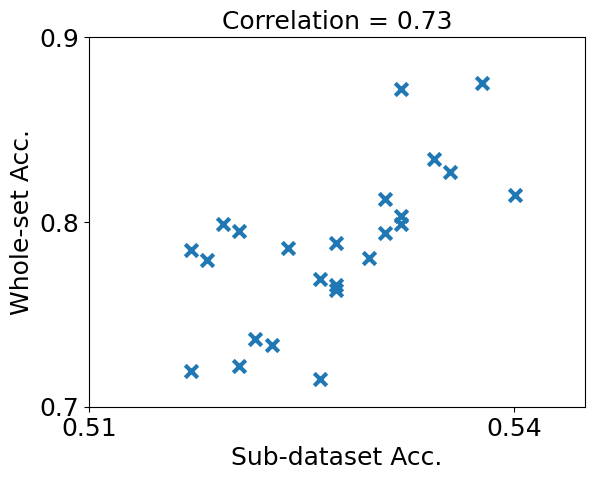}}
\subfigure[Random] {\includegraphics[width=1.75in]{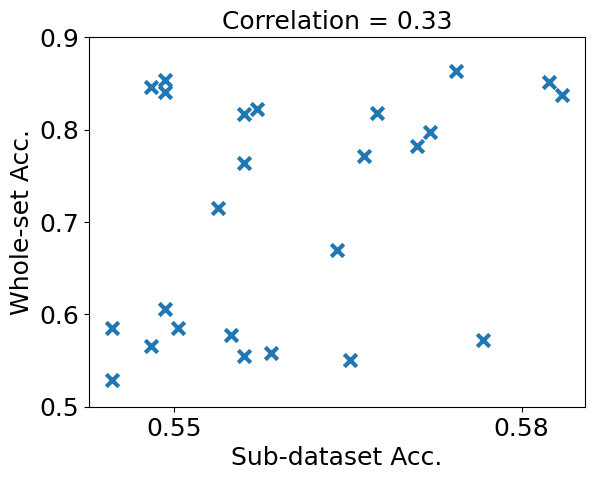}}
\subfigure[Herding] {\includegraphics[width=1.75in]{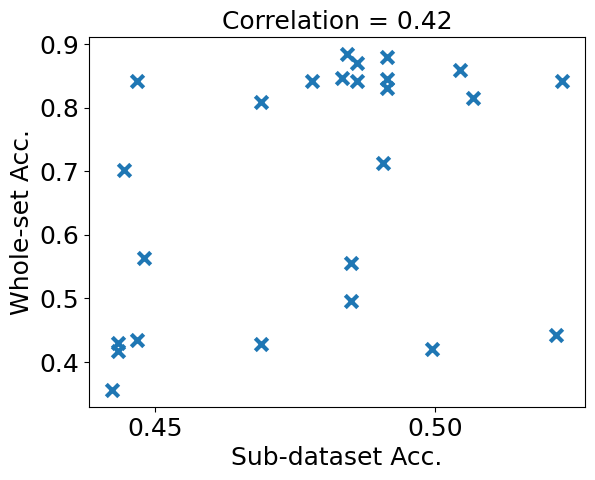}}
\subfigure[Early-stopping] {\includegraphics[width=1.75in]{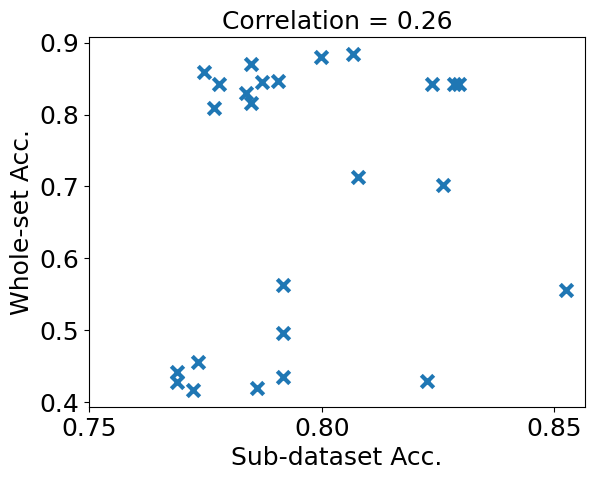}}
\caption{The performance of DDFAD in NAS.}
\label{fig:nas}
\end{figure*}

\subsection{Continual Learning}
\label{sec:Continual Learning}

We apply our DDFAD to a Class Incremental Continual Learning (CICL) task, where the objective of CICL is to learn a new class while preserving the performance in old classes. Following the SOTA continual learning baseline EEIL \cite{castro2018end}, we construct a limited budget rehearsal memory comprising representative samples from old classes. The function of this memory is to alleviate the catastrophic forgetting problem of DNNs. In our experiments, we replace the representative sample selection strategy in EEIL (i.e. herding), with our DDFAD and random selection and keep the rest the same to perform the CICL task. The representative samples pool is set to 20 audio clips for each old class.

As illustrated in Table \ref{table:The performance of DDFAD in class incremental learning}, DDFAD outperforms EEIL and random selection in the CICL task. This suggests that the memory of old classes constructed by our DDFAD contains more informative data for model training compared to EEIL and random selection methods. This improvement stems from the ability of DDFAD to distill knowledge from old classes more effectively, thereby aiding in mitigating catastrophic forgetting in CICL scenarios.

\begin{table}[ht]\small
\centering
\caption{The performance of DDFAD in class incremental continual learning (\%).}
\label{table:The performance of DDFAD in class incremental learning}
\resizebox{0.98\linewidth}{!}{
\begin{threeparttable}
\begin{tabular}{cc|ccc}
  \toprule
  \multirow{2}{*}{Dataset} & \multirow{2}{*}{Architecture} & \multicolumn{3}{c}{CICL methods} \\
    \cline{3-5} 
  &  &DDFAD   & EEIL \cite{castro2018end} &   Random \cite{rebuffi2017icarl}    \\
  \toprule
  \multirow{3}{*}{FSDD}&  ResNet18 &  	92.05&	80.51&	87.10 \\
  &  ConvNet &  92.92 & 76.41 & 83.33 \\
 &  VGG11 &  	91.67& 74.15&	80.77  \\
    \hline  
\multirow{3}{*}{UrbanSound}&  ResNet18 &  	46.25& 	21.58&	38.55  \\
  &  ConvNet &  	51.00& 	13.62&	15.22 \\
 &  VGG11 &  	47.12&	26.74&	31.74  \\
    \hline  
 \multirow{3}{*}{RAVDESS}&  ResNet18 &  	39.31&	15.94&	22.54 \\
  &  ConvNet &  	34.03&	13.19&	11.11  \\
 &  VGG11 &  	36.25&	16.72&	28.31  \\
    \bottomrule
\end{tabular}
    \end{threeparttable}}
\end{table}

\subsection{Neural Architecture Search}
\label{sec:Neural Architecture Search}

Our proposed DDFAD can also be applied to neural architecture search (NAS), which searches for the best network architecture for a given dataset. It typically needs significant training efforts of many candidate neural network architectures on the given dataset. Benefiting from the small size of the distilled dataset, it can be served as a sub-dataset to accelerate model evaluation in NAS.

Specifically, we follow the previous work \cite{zhao2020dataset} to design a set of candidate neural network architectures based on the considered ConvNet. We vary the depth, width, pooling, activation, and normalization layers of the ConvNet, producing 720 candidate architectures. These models are trained on the entire FSDD training dataset to establish ground-truth performance. Four NAS methods are considered for comparison, including random selection, herding selection, early-stopping and DDFAD. In terms of random selection, herding selection and DDFAD, we generate three sub-datasets using these methods with 20 audio clips per class. The models are trained for 100 epochs. For early-stopping, we train the model on the entire original training dataset for 10 epochs. Finally, we identify the top-performing architectures of these NAS methods and report their testing performance.

We illustrate the distribution of correlation between the NAS-based performance and the ground-truth performance on 5\% top-performing architectures in Figure \ref{fig:nas}. It can be seen that the NAS-based performance by DDFAD achieves the highest correlation (0.73) with the ground-truth performance. It demonstrates the promising applications of our DDFAD in NAS.

In addition to continual learning and neural architecture search, DDFAD can also be utilized to protect the privacy of the training dataset. Because the original training data is hard to recover from the distilled data, even in the case of training data leakage and member inference attacks \cite{bertran2024scalable}. The evaluations of data privacy protection are left for future work.




\section{Conclusions and Future Works}
\label{sec:Conclusions}
In this work, we propose a dataset distillation framework for audio data (DDFAD). Specifically, we propose the Fused Differential MFCC (FD-MFCC) as extracted features for audio data. It fuses the features of MFCC, the first-order difference of MFCC and the second-order difference of MFCC, which is more informative in the case of the small-scale distilled dataset. After that, we employ the matching training trajectory (MTT) distillation method to distill the FD-MFCC features. Finally, we propose an audio signal reconstruction algorithm based on the Griffin-Lim to rebuild the audio signal from the distilled FD-MFCC. Extensive experiments demonstrate the effectiveness and promising application prospects of DDFAD. 

For future works, we intend to explore other potential applications of audio dataset distillation, such as data privacy protection.  In addition, we aim to investigate dataset distillation methods under other tasks, such as object detection and natural language processing.


\bibliographystyle{IEEEtran}
\bibliography{distillation_for_voice}




\end{document}